\documentclass[aps,prc,showpacs,twocolumn,amssymb,floatfix]{revtex4}
\newcommand{\p}{\partial}

\newcommand{\ppkz}{\frac{\partial}{\partial k_z}}

\usepackage{graphicx}

\begin{document}

\title{Enhanced heavy  quark-pair production in strong SU(2) color field}
\author{P\'eter L\'evai}
\affiliation {KFKI RMKI Research Institute for Particle and Nuclear Physics, \\
P.O. Box 49, Budapest 1525, Hungary}

\author{Vladimir Skokov}
\affiliation {
Gesellschaft
f\"ur Schwerionenforschung mbH, Planckstr. 1,
D-64291 Darmstadt, Germany\\
and \\
Bogoliubov Laboratory of Theoretical Physics, 
Joint Institute for Nuclear Research, 
Dubna, 141980, Russia }

\date{September 12, 2009}

\begin{abstract}

Non-perturbative  charm and bottom quark-pair production is investigated in
the early stage of heavy ion collisions.
Following our earlier works, the time-dependent study is based on 
a  kinetic description 
of  fermion-pair  production in strong non-Abelian fields.
We introduce time-dependent chromo-electric external field
with a pulse-like time evolution, which
simulates the overlap of two colliding heavy ions. 
The calculations is performed
in a SU(2) color model with finite current quark masses. 
Yields of heavy quark-pairs
are compared to the ones of
light and strange quark-pairs.
We  show   that the small inverse duration time  of the field pulse 
determines the efficiency of the quark-pair production. Thus we do not see
the expected  suppression for heavy quark production, as follows from  
the   Schwinger formula for constant field, but rather 
an enhanced heavy   quark production 
at ultrarelativistic energies.   
We convert  pulse duration time-dependent results into
collisional energy dependence and introduce flavour-dependent  
energy  string tensions, which can be used in 
string based model calculations at RHIC and LHC energies.
\end{abstract}


\maketitle

\section{Introduction}

The main aim of ultrarelativistic heavy ion collisions is to create
extreme high energy densities and study the deconfinement
phase transition 
of colored quarks and gluons. Experiments at the BNL Relativistic Heavy Ion
Collider (RHIC) have been investigated the center of mass colliding region up to  
$\sqrt{s} = 200$ AGeV 
and detectors at CERN Large Hadronic Collider (LHC) are ready to
explore the energy range to $\sqrt{s} = 5500$ AGeV.
At such  high energies the colliding nuclei  are two colliding sheets
of nucleons with a huge Lorentz-contraction 
($\gamma_{cm} = 100$ at RHIC and $\gamma_{cm} = 2750$ at LHC), surrounded by a gluon cloud. 
Their overlap
results in a strong chromo-electric and 
chromo-magnetic field to be built up. Particles, namely gluons
and quark-antiquark pairs are  produced from this strong field, 
similarly to the Schwinger mechanism in quantum electrodynamics (QED)~\cite{schw51}.
The particle production rate  depends  on the
field strength, which  is varying in time.
The soft particles produced
in such a non-perturbative way form the bulk of the
wanted quark-gluon plasma,
after they successfully thermalized. Light and strange quarks
loose most of their original properties during thermalization,
but charm and bottom quarks conserve certain characteristic properties,
which can be studied after the whole evolution of the heavy ion 
collisions. In this paper we  investigate
the primordial non-perturbative production of heavy charm and bottom quarks, 
and explore the early stage of heavy ion collisions.

Recently the study of heavy quark production has received wide interest,
because open charm has been measured at RHIC in d+Au~\cite{Adams:2004fc}, 
Cu+Cu~\cite{Baumgart:2008xe},
and Au+Au~\cite{Abelev:2008hja} collisions. The
STAR~\cite{Adams:2004fc,Baumgart:2008xe,Abelev:2008hja}
and
PHENIX~\cite{Adare:2006hc,Adare:2006nq}
experiments 
have obtained different results (within a factor of 3), which finding opened 
vivid experimental and theoretical discussions~\cite{QM08}. 
A review on heavy-flavour
production has been published recently~\cite{Frawley:2008kk}. Theoretical calculations
based on perturbative quantum chromodynamics (pQCD) at fixed order
next to-leading logaritms (FONLL) have found that the
measured total charm cross section only comparable with the upper 
limit of the FONLL calculations. Thus there is  a room for non-perturbative
production channels. However, the situation is very complex,
as it was discussed in a recent paper~\cite{Pop:2009sd}: 		   
measured
heavy quark radiative and collisional energy loss in heavy ion
collisions must fit into
the picture, as well as the regeneration of charm hadrons in quark
coalescence  channels. 
The interest on this question supports the importance of
our investigation of primordial non-perturbative   heavy quark production.

Theoretical descriptions of particle production in high energy $pp$
collisions are based on the introduction of chromoelectric flux tube
('string') models~\cite{FRITIOF,HIJ,RQMD,QGSM,HIJINGBB}. 
String picture is a good example of how to convert the kinetic
energy of a collision into field energy, than later on gain 
the stored kinetic energy back.
However, at RHIC and LHC energies the string density is expected to be
so large that a strong collective gluon field will be formed in the whole
available transverse volume.
Furthermore, the gluon number will be so high that a classical gluon field
as the expectation value of the quantum field can be considered
in the reaction volume~\cite{Gyul97, Magas2002,Topor05}.
Alternatively at extremely high energies,  nucleus nucleus collisions 
can be described as two colliding sheets of Colored Glass Condensate. 
In the framework of this model it was shown that in the  early stage of 
collision  longitudinal color-electric and color-magnetic 
fields are created~\cite{Lappi:2006fp}. 
The properties of such non-Abelian classical fields and details
of gluon production
were studied very intensively during the last years, especially
asymptotic solutions (for summaries see Refs.~\cite{McLerran:2001sr,IanVen03}).
Fermion production was calculated recently~\cite{Gelis:2006yv,Gelis:2006cr,Blaizot:2004wv}.

Fermion pair production together with boson pair production
were investigated by different  models of particle
production from strong
Abelian~\cite{Gatoff87,Kluger91,Gatoff92,Wong95,Eisenberg95,Vin02,Per03,Prozorkevich:2004yp,Pervushin:2006vh,Mihaila:2009ge, Dawson:2009cn}
and non-Abelian~\cite{Prozor03,Diet03,HeinzOchs} fields.
These calculations concentrated
mostly on the bulk properties of the gluon and quark matter,
the time evolution of the system,
the time dependence of energy and particle number densities,
and the appearance of fast thermalization.

In our previous papers (see Ref.~\cite{SkokLev05,SkokLev07})
we investigated massless fermion and boson production in strong Abelian and
non-Abelian external electric field. During these calculations we
have realized, that the role of mass becomes  unimportant  when the
collisional energy is increasing and the the pulse duration time 
becomes comparable to the inverse quark mass~\cite{Levai:2008wf}.
In this paper 
we describe strange, charm, and bottom quark-pair production.
Motivated by  the problems raised  in Ref.~\cite{Pop:2009sd} we
investigate the role of string tension in the
Schwinger mechanism for heavy quark pair production. 

The energy dependece of the string tension was investigated earlier~\cite{Magas2002}. 
Instead of the usual $\kappa \sim 1$ GeV/fm value the much higher effective string tension, 
$\kappa\sim5-12$ GeV/fm appeared in the calculations. This open question 
motivates our investigation also. 

In this paper we solve the kinetic
model in the presence of an external SU(2) non-Abelian color field.
We focus on   
particle production with finite mass at different 
duration time of the quickly changing external field. 
Section 2 summarizes theoretical background of the kinetic equation
for color Wigner function. In Section 3 we 
consider the kinetic equation for pure longitudinal time-dependent 
SU(2) color field, which will be solved numerically. 
In Section 4 we summarize and discuss our numerical results
from general point of view. We discuss the collision energy 
and pulse duration time-dependence of heavy quark production. In a phenomenological way 
we introduce  flavour specific string tensions
to connect our numerical results and Schwinger estimate.

In Appendix A we derive general kinetic equation for 
the Wigner function starting from QCD Lagrangian.
In Appendix B we show exact solution of the kinetic equation 
for SU(2) color case, that supports our numerical calculation. 

\newpage 

\section{The kinetic equation for the Wigner function}

The equation of motion for color Wigner function 
$W(\mathbf{k},t)$ in the gradient  approximation reads (see Appendix A for details)
\begin{eqnarray}
&&\partial_t W+ \frac{g}{8}\frac{\partial}{\partial k_i}
\left( 4\{W,F_{0i}\} 
+ \right. \nonumber 
\\
&&\left. 
+2\left\{F_{i\nu},[W,\gamma^0 \gamma^\nu]\right\}-
\left[F_{i\nu},\{W,\gamma^0 \gamma^\nu\}\right] \right)=\nonumber\\
&&=ik_i \{\gamma^0 \gamma^i,W\}-im[\gamma^0,W] +ig \left[A_i\,, 
[\gamma^0 \gamma^i ,W]\right] . \ \ \ 
\label{KEW}
\end{eqnarray}
Here $m$ denotes the current mass of the fermions, 
$g$ is the coupling constant, $A_\mu$ is the 4-potential of an
external space-homogeneous color field and $F_{\mu\nu}$ is corresponding field tensor 
\begin{equation}
F_{\mu \nu} = \partial_\mu A_\nu - \partial_\nu A_\mu - i g  [ A_\mu, A_\nu ].
\end{equation}
The validity of gradient approximation requires that the Wigner function is sufficiently smooth 
in momentum space and the field strength varies slowly in coordinate space. The 
corresponding characteristic lengths must satisfy the following 
relation $(\Delta p)_W (\Delta x)_F \gg 1$, 
where $(\Delta p)_W$ is connected to the momentum gradients of the Wigner function
and $(\Delta x)_F$ to space  gradients  of the field.   

The color decomposition of the Wigner function with SU($N_c$) 
generators in the fundamental representation is given by 
\begin{equation}
W = W^s + W^a t^a, \,\, \ \ \ a = 1,2,..., N_c^2-1 \ ,
\label{color_decomposition}
\end{equation}
where $W^s$ is the color singlet part and $W^a$ is the color multiplet components. 
It is also convenient to perform spinor decomposition separating scalar $a$, vector
$b_\mu$, tensor $c_{\mu\nu}$, axial vector $d_\mu$ and pseudo-scalar parts $e$:
\begin{equation}
W^{s|a} = a^{s|a} + b^{s|a}_\mu \gamma^\mu + c^{s|a}_{\mu\nu} \sigma^{\mu\nu} +
d^{s|a}_\mu \gamma^\mu \gamma^5 + i e^{s|a} \gamma^5.
\label{Clifford_decomposition}
\end{equation}
The asymmetric tensor  components  of the Wigner function is convenient to 
decompose into axial and polar vectors $c_1^j=c^{j0}$ and 
$c_2^j=\frac{1}{2}\epsilon^{0 \omega\rho j}c_{\omega\rho}$ correspondingly. 

\section{Kinetic equation in SU(2) with color isotropic external field}
After the color and spinor decomposition of the equations for the
Wigner function in case of 
pure longitudinal external SU(2) color field with fixed color direction 
$A^a_z = A^\diamond_z n^a$, where  $n^a n^a=3$ and $\p_t n^a=0$~\cite{SkokLev07}, 
we obtain the following  system of equations for  the singlet component
\begin{eqnarray}
&&\p_t a^s + \frac{3 g}{4} E^\diamond_z  \ppkz  a^\diamond  = - 4 \mathbf{k} \mathbf{c}_1^s,    \\
&&\p_t e^s + \frac{3 g}{4} E^\diamond_z  \ppkz    e^\diamond  = - 4 \mathbf{k} \mathbf{c}_2^s - 2  m d^s_0,\\
&&\p_t b^s_0 + \frac{3 g}{4} E^\diamond_z   \ppkz  b^\diamond_0 =  0,\\
&&\p_t \mathbf{b}^{s} + \frac{3g}{4} E^\diamond_z \ppkz \mathbf{b}^\diamond =  2   [ \mathbf{k} \times \mathbf{d}^s ]  + 4  m \mathbf{c}_1^s,\\
&&\p_t d^s_0 + \frac{3g}{4} E^\diamond_z \ppkz   d^\diamond_0  =  2  m  e^s, \\
&&\p_t \mathbf{d}^s + \frac{3g}{4} E^\diamond_z \ppkz   \mathbf{d}^\diamond =  2 [\mathbf{k} \times  \mathbf{b^s}], \\
&&\p_t \mathbf{c}_1^s  + \frac{3g}{4} E^\diamond_z \ppkz \mathbf{c}_1^\diamond  = a^s \mathbf{k} - m \mathbf{b}^s,\\
&&\p_t \mathbf{c}_2^s  + \frac{3g}{4} E^\diamond_z \ppkz \mathbf{c}_2^\diamond  =  e^s \mathbf{k};
\end{eqnarray}
and the triplet  components 
\begin{eqnarray}
&&\p_t a^\diamond + g E^\diamond_z \ppkz  a^s   = - 4 \mathbf{k} \mathbf{c}_1^\diamond, \\
&&\p_t e^\diamond + g E^\diamond_z \ppkz e^s  = - 4 \mathbf{k} \mathbf{c}_2^\diamond- 2m d^\diamond_0, \\
&&\p_t b^\diamond_0 + g E^\diamond_z \ppkz  b^s_0  = 0,\\
&&\p_t \mathbf{b}^\diamond + g E^\diamond_z \ppkz \mathbf{b}^s 
=  2 [ \mathbf{k} \times  \mathbf{d}^\diamond ]  + 4 m \mathbf{c}_1^\diamond,\\
&&\p_t d^\diamond_0 + g E^\diamond_z \ppkz  d^s_0 \delta^{b c}  
= 2 m e^\diamond,\\
&&\p_t \mathbf{d}^\diamond + g E^\diamond_z \ppkz \mathbf{d}^s 
= 2 [ \mathbf{k} \times \mathbf{b}^\diamond ],\\
&&\p_t \mathbf{c}_1^\diamond + g E^\diamond_z \ppkz \mathbf{c}_1^s
= a^\diamond \mathbf{k}  - m \mathbf{b}^\diamond,\\
&&\p_t \mathbf{c}_2^\diamond + g E^\diamond_z \ppkz \mathbf{c}_2^s 
= e^c \mathbf{k}.
\end{eqnarray}

The distribution function for massive fermions is completely  defined by components 
${a,b}$~\cite{SkokLev07}:
\begin{equation}
f_q(\mathbf{k},t) = \frac{m a^s(\mathbf{k},t) +  \mathbf{k} \, \mathbf{b}^s(\mathbf{k},t)}
{\omega(\mathbf{k})}  + \frac{1}{2}, 
\label{DF}
\end{equation}
where ${\omega(\mathbf{k})} = \sqrt{\mathbf{k}^2+m^2}$. 
Thus for time- and momentum-dependent distribution 
functions scalar $a$,  vector $b_\mu$, axial vector $d_\mu$, and tensor $c$  components   
of the Wigner  function are needed, only. 

The initial conditions for the Wigner function in vacuum reads (see Appendix A)
\begin{eqnarray}
a^s &=& -\frac{1}{2} \frac{m}{\omega}, \\
\mathbf{b}^s &=&  -\frac{1}{2} \frac{\mathbf{k}}{\omega}.
\label{Initial}
\end{eqnarray}

Considering vacuum initial condition symmetry 
we obtain the following equations for
the singlet (we redefined $\mathbf{c}=\mathbf{c}_1$ to simplify reading)
\begin{eqnarray}
\label{final_bs}
&&\p_t a^s + \frac{3 g}{4} E^\diamond_z  \ppkz  a^\diamond  = - 4 (k_z c_z^s + k_\perp c_\perp^s),    \\
&&\p_t b_z^s + \frac{3g}{4} E^\diamond_z \ppkz b_z^\diamond = 2 k_\perp  d_x^s   +  4  m c_z^s,\\
&&\p_t b_\perp^s + \frac{3g}{4} E^\diamond_z \ppkz b_\perp^\diamond = - 2 k_z  d_x^s   +  4  m c_\perp^s,\\
&&\p_t d_x^s + \frac{3g}{4} E^\diamond_z \ppkz d_x^\diamond = 2( k_z  b_\perp^s - k_\perp b_z^s ),\\
&&\p_t c_z^s  + \frac{3g}{4} E^\diamond_z \ppkz c_z^\diamond  = a^s k_z - m b_z^s,\\
&&\p_t c_\perp^s  + \frac{3g}{4} E^\diamond_z \ppkz c_\perp^\diamond  = a^s k_\perp - m b_\perp^s \ ;
\label{final_es}
\end{eqnarray}
and for the triplet  components 
\begin{eqnarray}
\label{final_bm}
&&\p_t a^\diamond + g E^\diamond_z \ppkz  a^s   = - 4 (k_z c_z^\diamond + k_\perp c_\perp^\diamond), \\
&&\p_t b_z^\diamond + g  E^\diamond_z \ppkz b_z^s = 2 k_\perp  d_x^\diamond   +  4  m c_z^\diamond,\\
&&\p_t b_\perp^\diamond + g E^\diamond_z \ppkz b_\perp^s = - 2 k_z  d_x^\diamond   +  4  m c_\perp^\diamond,\\
&&\p_t d_x^\diamond + g  E^\diamond_z \ppkz d_x^s = 2( k_z  b_\perp^\diamond - k_\perp b_z^\diamond ),\\
&&\p_t c_z^\diamond  +  g E^\diamond_z \ppkz c_z^s  = a^\diamond k_z - m b_z^\diamond,\\
&&\p_t c_\perp^\diamond  + g E^\diamond_z \ppkz c_\perp^s  = a^\diamond k_\perp - m b_\perp^\diamond \ ; 
\label{final_em}
\end{eqnarray}
where we introduced the following vector decomposition 
\begin{equation}
\mathbf{v} = v_z \mathbf{n} + v_\perp \frac{\mathbf{k}_\perp}{k_\perp} 
+ v_x \left[ \mathbf{n} \times \frac{\mathbf{k}_\perp}{k_\perp} \right] \ . 
\end{equation}
Here  the unit vector collinear to the field direction, $\mathbf{n}$,  is given by 
$\mathbf{n} = \mathbf{E}^\diamond/\vert \mathbf{E}^\diamond \vert = (0,0,1)$ 
and $\mathbf{k}_\perp = (k_1,k_2,0)$.
As follows from Eqs. (\ref{final_bs}-\ref{final_em}), 
the axial part of vector $b_x$,  tensor components $c_x$, longitudinal $d_z$ and perpendicular $d_\perp$
parts of axial vector components do not contribute to the evolution of the distribution function. 

These equations  could be further simplified and transformed to ones that are similar to Abelian 
case (see Appendix B for details).

\section{Numerical results and discussions} 

\subsection{General results}

In Ref.~\cite{SkokLev07} we have solved the above equations 
for massless (light) quarks
and described  their longitudinal and transverse momentum distributions.   
Here we focus on the integrated particle
yields and discuss the obtained results, focusing 
on massive (heavy) quark production. 

In the numerical calculation we have used the following parameters: 
the maximal magnitude of the field  $E_0=0.68$ GeV/fm; the strong coupling constant $g=2$;
the current quark masses $m_{u,d}=8$ MeV, $m_s=150$ MeV, $m_c=1.2$ GeV,
$m_b=4.2$ GeV for light, strange, charm and bottom  quarks,
correspondingly. The value of maximal magnitude of the field corresponds 
to the effective string tension $\kappa\sim1.17$ GeV/fm. The reason why we use this value will be 
explained below,  in Subsection~\ref{Sdep}.   

The particle production is  ignited by a pulse-like color field
simulating a  heavy ion collision~\cite{SkokLev07}: 
\begin{equation}
E^\diamond(t) = E_0 \cdot \left[ 1- \textrm{tanh}^2 (t/\tau) \right], 
\label{field}
\end{equation}
where $\tau$ is a pulse duration time.

\begin{figure}[t]
\centerline{
\includegraphics[height=6cm] {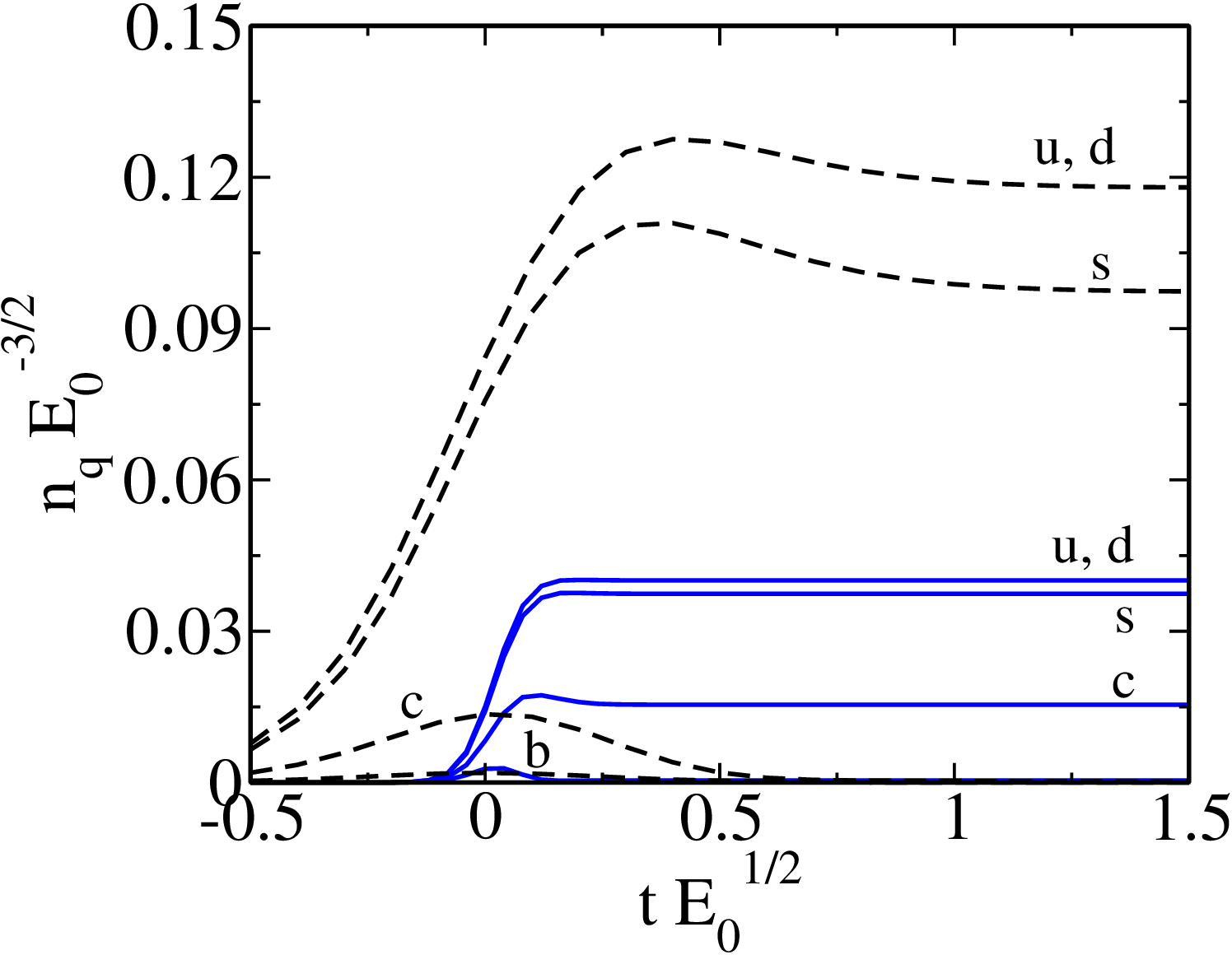}}
\caption{
\label{nt}
The dimensionless  quark number density  for different flavours, $n_q(t) E_0^{-3/2}$,
as a function of dimensionless time  $tE_0^{1/2}$ 
for different pulse duration, $\tau E_0^{1/2}=0.1$ (solid lines) and 
$\tau E_0^{1/2}=0.5$ (dashed lines).} 
\end{figure}

\begin{figure}[b]
\centerline{
\includegraphics[height=6cm] {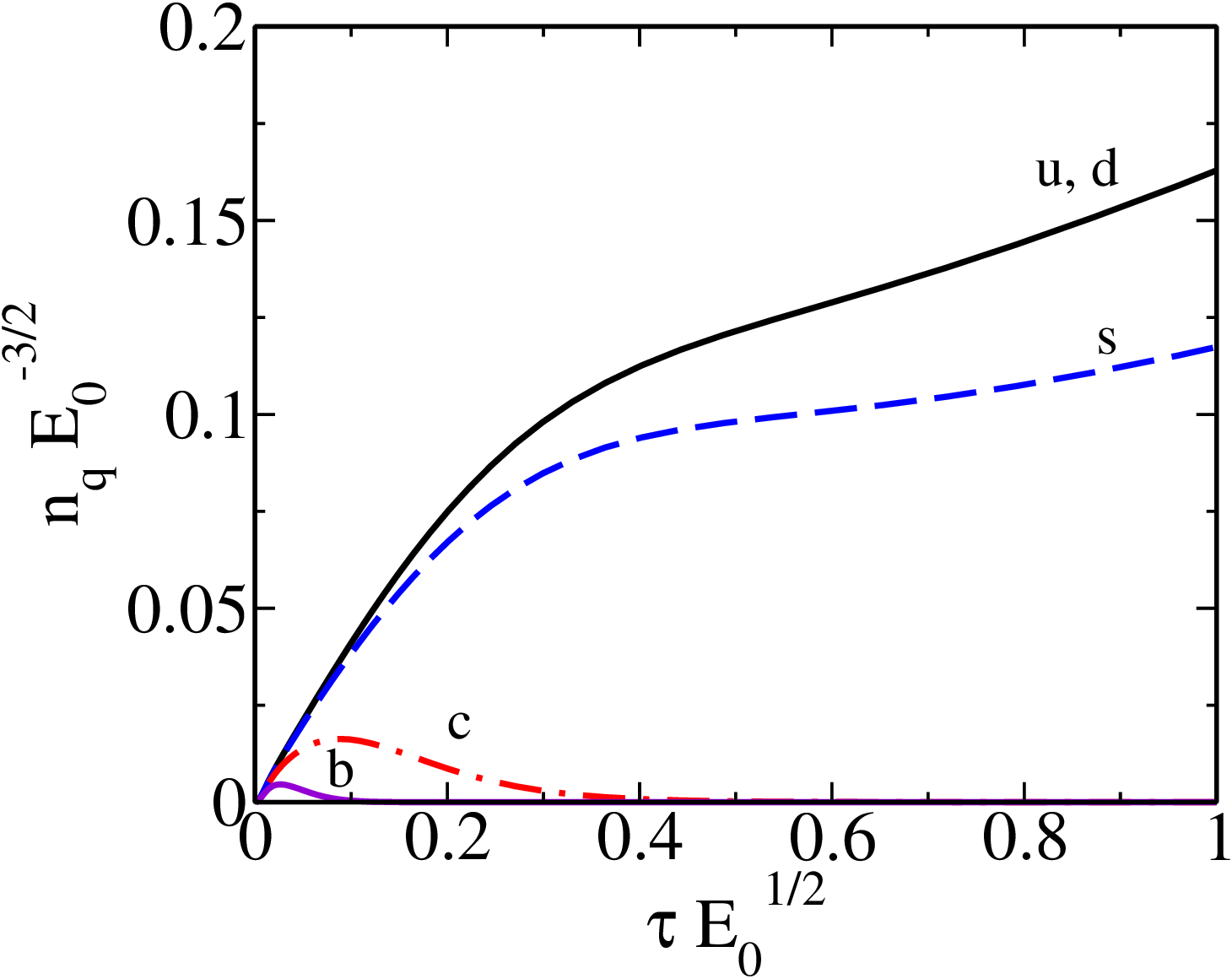}}
\caption{
The dimensionless  quark number densities  at the asymptotic final state,
$n_q(t \gg \tau) E_0^{-3/2}$,  
as a function of pulse duration time  $\tau E_0^{1/2}$. 
}
\label{n}
\end{figure}

The suppression factor of  heavier quark $Q$ to light one $u$  is defined 
in the asymptotic future (c.f. ~\cite{schw51}), $t \gg \tau$, as 
\begin{equation}
\gamma^Q = \lim_{t\to\infty} n_Q(t)/n_{u}(t). 
\end{equation}
Here $n_q(t)$ is the number density of corresponding quarks given by 
\begin{equation}
n_q(t) = 4 N_c\int \frac{d^3k}{(2\pi)^3} f_q(\mathbf{k},t) \ ,  
\label{numberdens}
\end{equation}
where $q$ denotes different quark flavours, $q$=u, d, s,  c, b.

\begin{figure}[t]
\centerline{
\includegraphics[height=6cm] {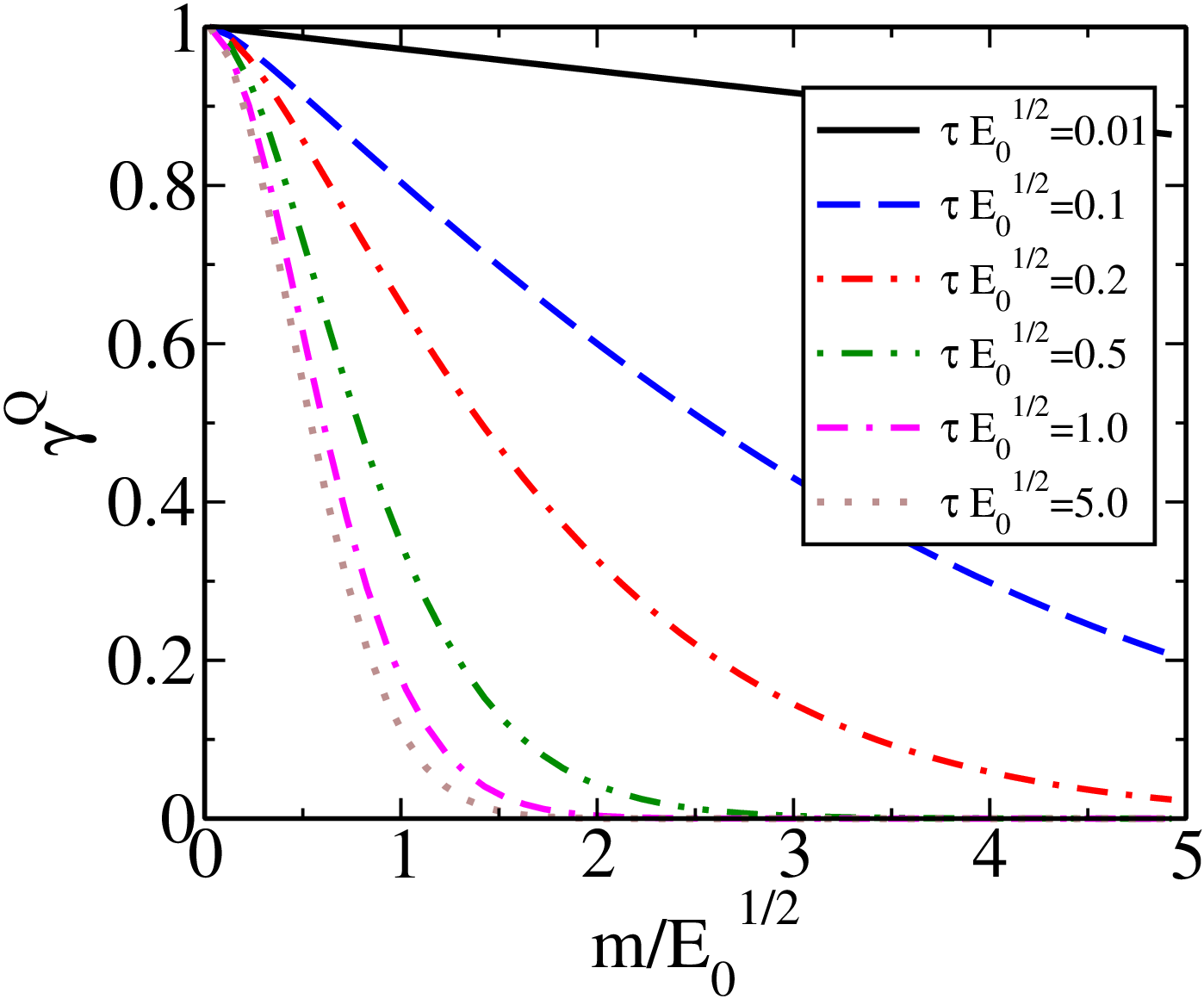}}
\caption{
The suppression factor,   $\gamma^Q$, at different
pulse duration time as a function of the dimensionless mass parameter, 
$m/E_0^{1/2}$. }
\label{mdep}
\end{figure}

\begin{figure}[b]
\centerline{
\includegraphics[height=6cm] {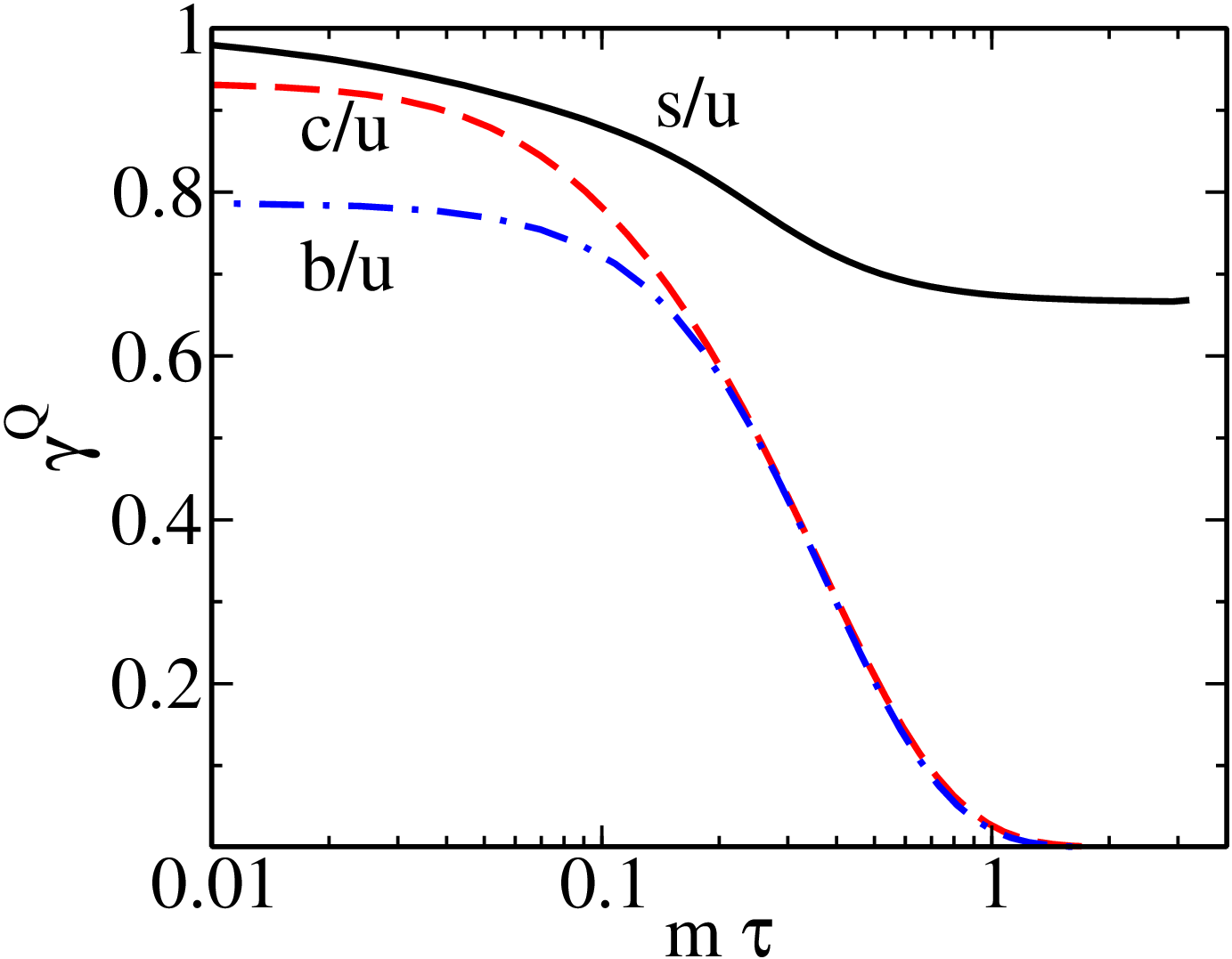}}
\caption{
The suppression factor for heavy flavours at different values 
of $m\tau$. }
\label{suponmt}
\end{figure}

In Fig. \ref{nt} the time evolution of quark number densities, $n_q$, are displayed
for different pulse  duration time,
$\tau E_0^{1/2}=0.1$ and 0.5. 
For short pulse  the quark  number  densities are comparable with each other (solid lines). 
In this case the particle production happens during the whole evolution of the field.
In contrast to this,
for long pulse, the number of produced charm and bottom 
quarks becomes negligible in the
final state, because their production is balanced by annihilation. In U(1) color 
case the annihilation
term can be identified clearly, see Ref.~\cite{SkokLev05}.

This dependence on the pulse duration time  is also demonstrated in Fig.~\ref{n}. 
This figure clearly displays that both charm and bottom  quark production 
is  substantially enhanced in the cases of short pulse.
This enhancement has a maximum at $\tau \sim 0.1\sqrt{E_0}$ for charm and 
at even smaller $\tau$ for heavier
bottom quark.
In opposite to the heavy  quarks, light and strange quark productions are
increasing with the pulse duration time without any local maximum.

Fig.~\ref{n} displays  an important result that at small value of 
$\tau$ the heavy quark production has a pronounced maximum with value 
well beyond the well known  asymptotic Schwinger estimate,  see Ref.~\cite{schw51}.

We further investigate  the suppression factor and its dependence on pulse duration 
time and quark masses.
Fig. \ref{mdep}  displays  
the suppression factor,
$\gamma^Q$,  for different pulse duration time, as a function of quark mass. 
For short pulse the suppression factor is
decreasing almost linearly with increasing quark mass value. For long  pulse 
we can see a very fast ($\sim \exp\{-m^2/E_0\}$) drop, which is consistent with
the Schwinger formula.

In Fig.~\ref{suponmt} we demonstrate the pulse duration time  dependence of the 
suppression factor for different flavours. As it can be seen the dramatic  
change of the suppression factors for heavy quarks happens in the region $0.1<m\tau<1$.  
Actually, as it is shown in Appendix B,  there are two dimensionless 
parameters that control the behaviour of the particle production. 
One is the dimensionless pulse duration, $m \tau$; the other is  
adiabaticity  parameter, $\Gamma_K\sim m/({E_0 \tau})$. The Schwinger 
formula is valid for the combination $m\tau\gg1$ and $\Gamma_K\ll1$. 

Note, that  at short pulse, the relative charm and bottom  production 
is surprisingly large, which does not follow any earlier
expectation.

\subsection{Energy and pulse duration time    dependence}
\label{Sdep}

To fix  free  parameters we use the  following  simple model. 
The field depends on two unknown parameters, $\tau$ 
and $E_0$. We will fix them as the best 
fit of the suppression factors for primordial strange and
charm quarks obtained in a quark coalescence calculation~\cite{Levai:2000ne,Levai:2008me} 
at RHIC energy, $\sqrt{s}=200 A$GeV. 
The suppression factors are $\gamma^s = 0.88$ and 
$\gamma^c = 6\cdot10^{-2}$. 
The best fit reads  
$E_0=0.68$ GeV/fm and $\tau_0=0.134$ fm$/c$.
Surprisingly, our simple model provides  reasonable  
values for these parameters. 

From  intuitive reasons the duration of field pulse is proportional 
to the time of two Lorenz-contracted heavy  ions pass each other at almost speed of light,
i.e. 
\begin{equation}
\tau \simeq \alpha \frac{2 R}{\gamma_{cm}},  
\label{taus}
\end{equation}
where $R$ is the radius of a nuclei, 
$\gamma_{cm}\simeq \sqrt{s}/(2 {\rm GeV})$ is the gamma-factor, 
$\alpha$ is an unknown  proportionality coefficient.
In case of gold-gold collision at RHIC energy
we obtained $\alpha=0.96$ from the best fit values given above.
We further assume that $\alpha$ weakly  depends  on the collision energy
and this dependence can  be neglected. Thus having 
the value of  $\alpha$ in hand we can transform 
the duration time dependence  to the 
collision energy dependence.  
Although this conversion  is oversimplified 
(e.g. it does  not take
into account any stopping effects), we expect 
to obtain results of  the right order of magnitude.  
The extracted  numbers make possible  to interpret  our numerical results
on realistic basis, namely energy scales.

\begin{figure}[t]
\centerline{
\includegraphics[height=6cm] {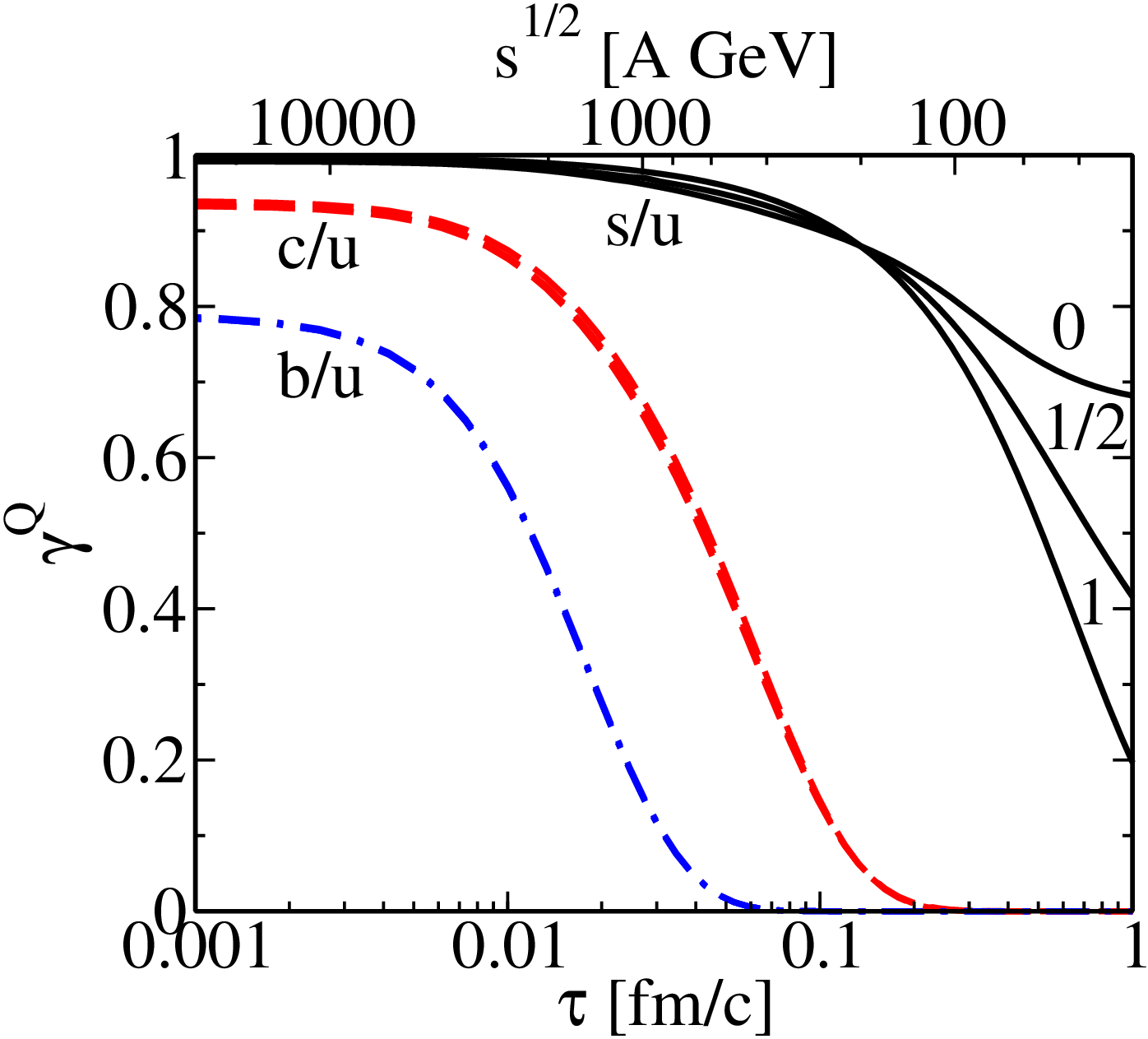} }
\caption{
The suppression factor, $\gamma^Q$, for strange and  heavy flavours as a function of 
pulse duration time and collision energy. The number marking the curves for 
strangeness suppression 
correspond to different values of $\beta$, see Eq.~(\ref{betas})     
}
\label{gamma}
\end{figure}

In  Fig.~\ref{gamma} we show   the time and energy dependence
of the suppression factor for strange, charm and bottom quarks.
To  demonstrate how robust our results we  calculated the 
suppression  factors assuming   
different  
pulse duration time (corresponded to different collision energies)
dependence  of the  field pulse magnitude $E_0(\tau)$.  
We consider three special cases 
\begin{equation}
E_0(\tau) = E_0 \cdot \left(\frac{\tau_0}{\tau}\right)^\beta, \quad \beta=0,\ 1/2,\ 1.
\label{betas}
\end{equation}
The constant $E_0(\tau)$ is recovered if $\beta=0$.
The second choice, $\beta=1/2$,  corresponds to 
finite number of quarks for short pulses,  $\tau\to0$,
(see Appendix B, Eq.~(\ref{shortpulse})).
Finally,  $\beta=1$ results in divergent number of quarks for  $\tau\to0$
(see Appendix B, Eq.~(\ref{fdiverg})).

In Fig.~\ref{gamma} we include these numbers, $\beta$, 
to distinguish  between different cases for the strange quark suppression factor.
For the heavy quarks all three curves are  indistinguishable. 
As it can be  seen,  for strange quark 
the difference between the above  cases is only important for low 
energy collisions, or long pulse duration.
At  constant $E_0$, and in the limit  $\tau\to\infty$  the suppression
factor for the strange quark production  tends to the Schwinger limit, $\sim 0.74$.
For charm  and bottom quarks the  Schwinger limit is negligibly small, see the suppression 
factor, $\gamma^Q$, for experimentally favoured energies in Table I.

\begin{table}
\begin{tabular}{ | c | c | c | c | c| c| c| }
\hline
&$\gamma_\infty$ & 130A GeV  &  200A GeV  & 1A TeV & 2A TeV & 5.5A TeV \\
\hline 
s  & 0.74 & 0.84  &  0.88  & 0.96 & 0.98 & 0.99 \\
\hline
c  & 3$\cdot 10^{-9}$ & 9$\cdot 10^{-3}$   & 0.06 & 0.66 & 0.82 & 0.91\\
\hline
b  & $\sim0$ & $\sim$0  & $10^{-6}$  & 0.15 & 0.45 & 0.72  \\
\hline
\end{tabular}
\caption{ The suppression factor, $\gamma^Q$,  for experimentally favoured
energies. The calculations are done with a string tension $\kappa\simeq1.17$, $\beta=0$. 
The result obtained  by Schwinger formula is denoted by $\gamma_\infty$. 
}
\end{table}

\subsection{Effective string tension}
In the previous subsections 
we have obtained  
that the time dependence of the strong color field
can induce an enhancement of heavy quark production 
(especially in a case of  pulse duration times  
corresponded to high energy collisions). 
In usual string-based models~\cite{FRITIOF,HIJ,RQMD,QGSM,HIJINGBB} 
these abundances can be introduced by increasing the effective string  
tension in the Schwinger formula.
However 
our results show that 
different effective string tensions should 
be introduced for  different quark flavours. Here we calculate these string tension values 
and offer them for later use.

\begin{figure}[b]
\centerline{
\includegraphics[height=6cm] {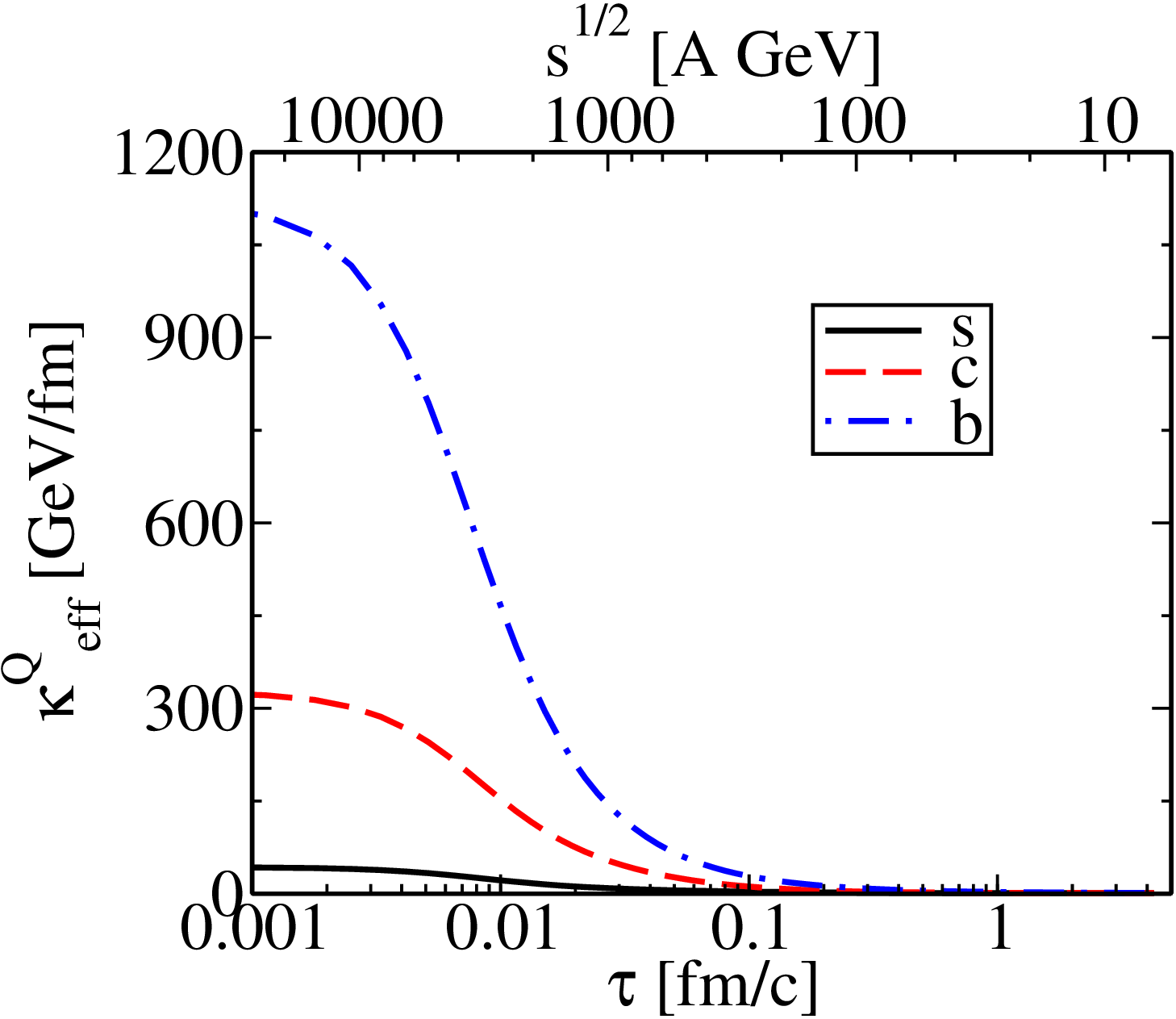}}
\caption{ 
The effective string tension, $\kappa^{Q}_{eff}$, defined by Eq.~(\ref{NC}) 
as a function of pulse duration time, $\tau$,  and collision energy, $\sqrt{s}$.
}
\label{effective_kappa_sqrt_s}
\end{figure}

Let us recall the Schwinger formula 
for particles with mass $m$
\begin{equation}
\frac{d N}{dt d^3x } = \frac{\kappa^2}{4\pi^3} \exp\left( - \frac{\pi m^2}{\kappa} \right).
\label{schwn}
\end{equation}
Here $\kappa$ is the string tension. According to this formula 
the suppression factor of heavier ($Q$) to light quarks ($q$)  is given
by 
\begin{equation}
\gamma^{Q}_{\infty}  = \exp \left( - \frac{\pi (m_Q^2-m_q^2)}{\kappa} \right).
\end{equation}
This formula is valid for the case of arbitrary N in SU(N), see Ref.~\cite{Gyulassy:1985}.

At first we extract an effective string tension 
assuming 
its common value for light and 
heavy quarks.
Providing our numerical calculation of suppression factor (see Fig.~\ref{gamma}) for 
Q-flavour, $\gamma^Q(\tau)$,  we solve the equation 
\begin{equation}
\gamma^Q_{\infty}(\kappa^Q_{ {\rm eff}}) = \gamma^Q(\tau)
\label{NC}
\end{equation}
to find the effective string tension dependence on pulse duration time, 
$\tau$, and subsequently  on the  collision energy, $\sqrt{s}$,  for given quark flavour.

\begin{figure}[b]
\centerline{
\includegraphics[height=6cm] {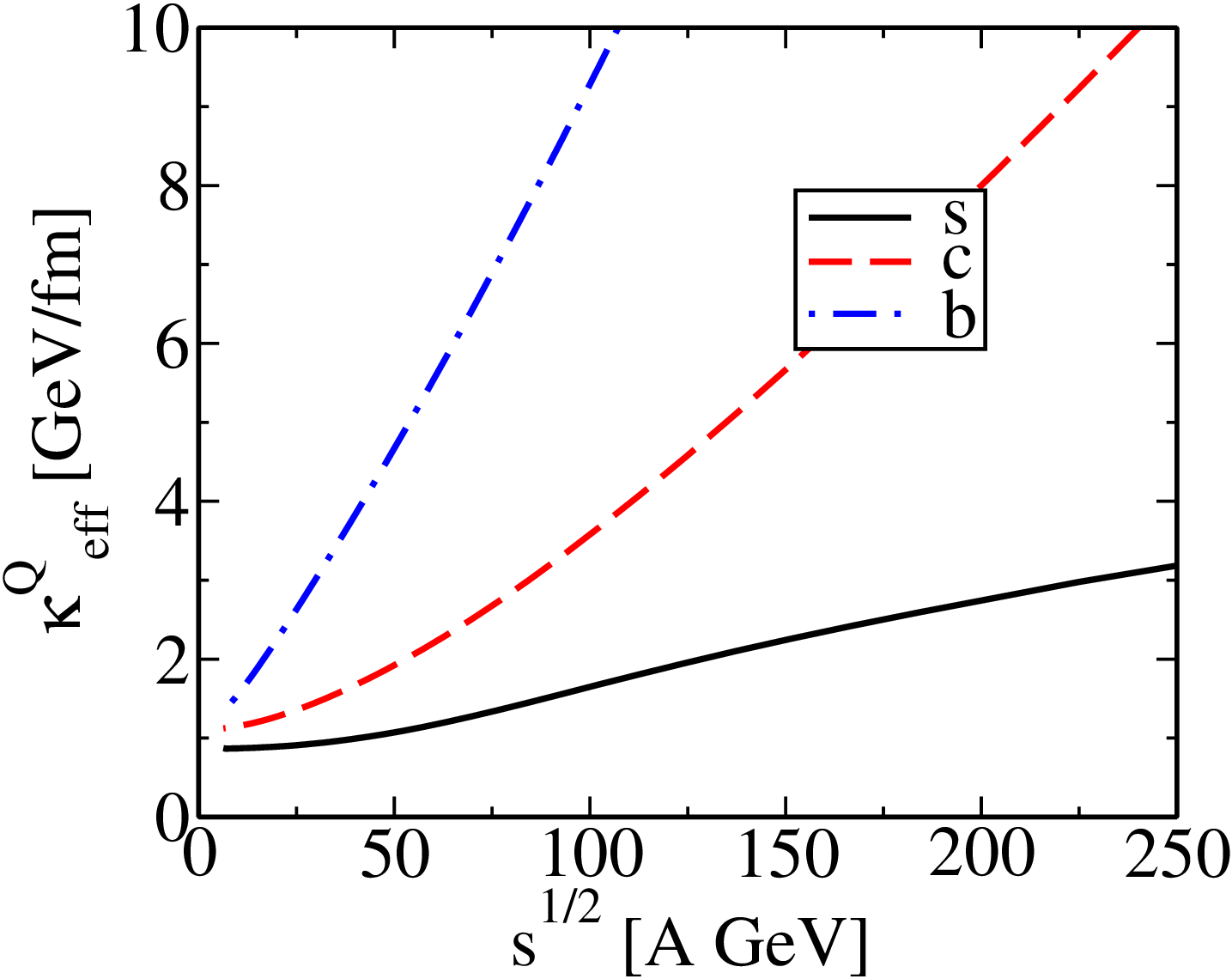}}
\caption{ The same as in Fig. \ref{effective_kappa_sqrt_s} in linear scales, 
but zoomed to  RHIC energy
range.
}
\label{effective_kappa_sqrt_s_zoomed}
\end{figure}

Fig.~\ref{effective_kappa_sqrt_s}  displays results in a logarithmic scale. 
As it can be seen, the values of $\kappa^Q_{ {\rm eff}}$ are very much different 
for strange, charm and bottom quarks. In Fig.~\ref{effective_kappa_sqrt_s_zoomed} 
we show our results  in a linear scale focusing on RHIC energy range. 
These large string tension values should be applied to light quark production as well, 
but we see three different values. 
This indicates the need of introduction of  effective string tension 
in a different way.

We keep the usual string tension for light quark $\hat{\kappa}^u_{ \rm eff}\simeq 1.17$ GeV/fm 
and introduce ``flavour specific'' effective  string tensions for heavier flavours. 
On the basis of Eq.(\ref{schwn}) we obtain the suppression factor 
\begin{equation}
\hat{\gamma}^Q_{\infty} = \left( \frac{\hat{\kappa}^Q_{\rm eff}   } {   \hat{\kappa}^u_{ \rm eff}   }
\right)^2
\exp\left( -\pi \frac{m_Q^2}{\hat{\kappa}^Q_{\rm eff}}  +  \pi \frac{m_u^2}{\hat{\kappa}^u_{\rm eff}}  \right).
\end{equation}

\begin{figure}[t]
\centerline{
\includegraphics[height=6cm] {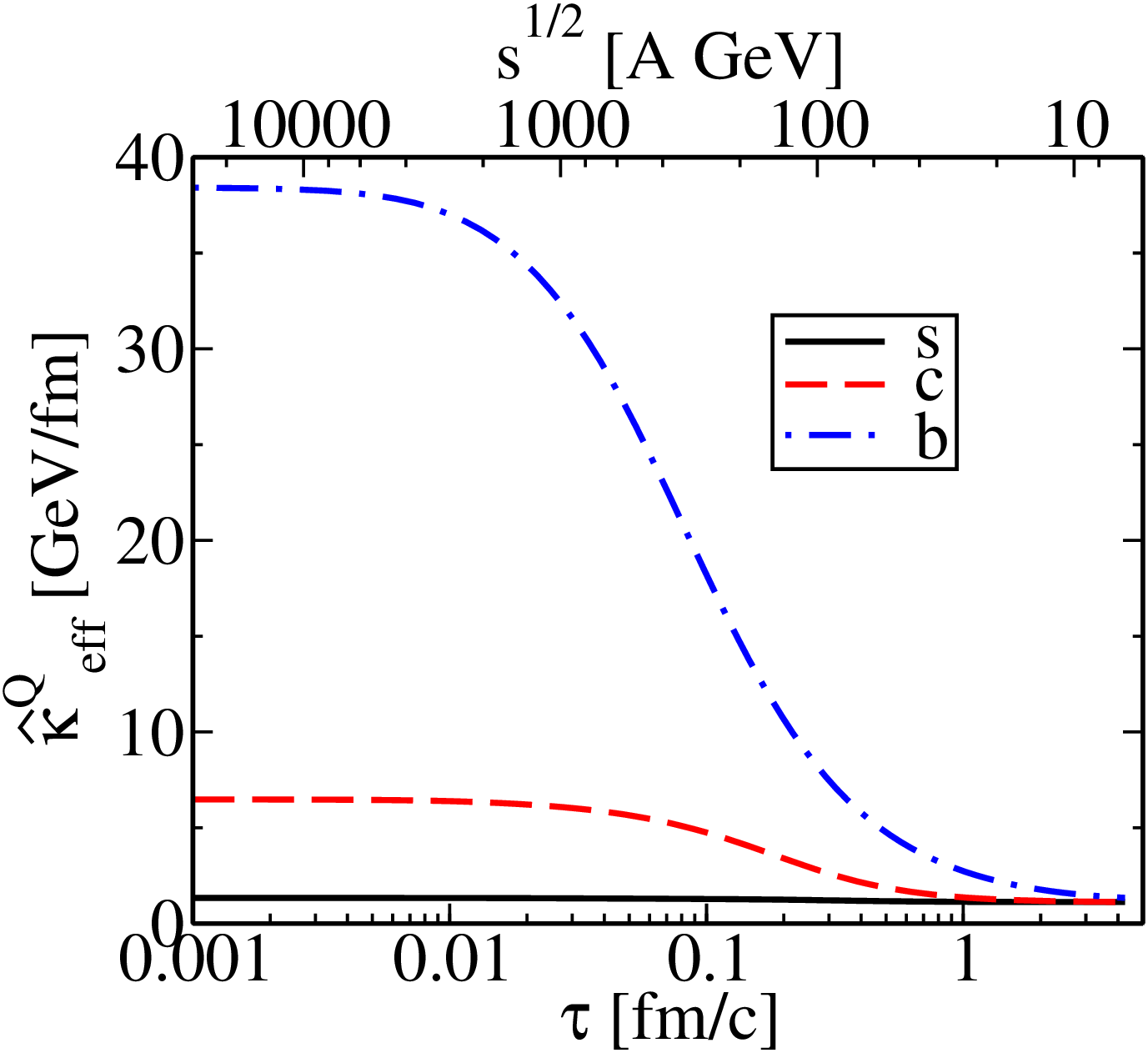}}
\caption{
The ``flavour specific'' effective string tension, $\hat{\kappa}^Q_{eff}$
as a function of  pulse duration time, $\tau$,  and collision energy, $\sqrt{s}$.
}
\label{effective_kappa_sqrt_s_nc}
\end{figure}

\begin{figure}[b]
\centerline{
\includegraphics[height=6cm] {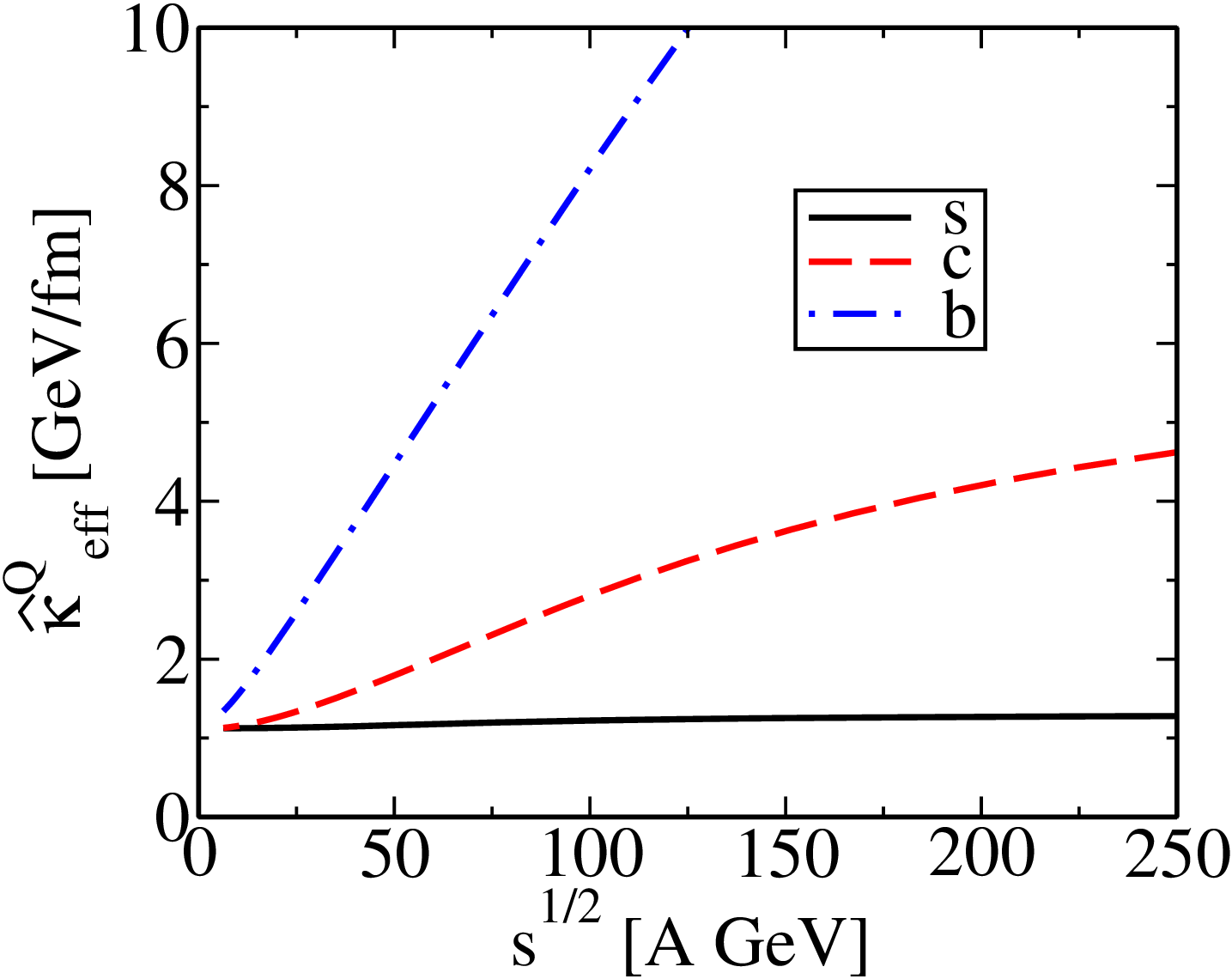}}
\caption{
The same as in Fig. \ref{effective_kappa_sqrt_s_nc} in linear scales, but zoomed to  RHIC energy
range.
}
\label{effective_kappa_sqrt_s_nc_zoomed}
\end{figure}

Extracting  such a   ``flavour specific''   effective string tensions from the numerically 
calculated values we obtain Fig.~\ref{effective_kappa_sqrt_s_nc} and 
Fig.~\ref{effective_kappa_sqrt_s_nc_zoomed}. We would like to emphasize 
the difference between $\kappa^Q_{\rm eff}$ and $\hat{\kappa}^Q_{\rm eff}$, 
which is demonstrated by their different values.  
For easier use we generated the Table II displaying the 
``flavour specific'' effective string tensions 
for experimentally favoured collision energies.  

\begin{table}
\begin{tabular}{ | c | c | c | c | c| c| }
\hline
& 130A GeV  &  200A GeV  & 1A TeV & 2A TeV & 5.5A TeV \\
\hline 
u, d   & 1.17  &  1.17  & 1.17 & 1.17 & 1.17 \\
\hline 
s  & 1.24  &  1.26  & 1.32 & 1.33 & 1.34 \\
\hline
c  & 3.32  &  4.2  & 6.1 & 6.3 & 6.5 \\
\hline
b  & 10.3  & 14.7  & 32 & 36 & 38\\
\hline
\end{tabular}
\caption{ The ``flavour specific'' effective string tension,
$\hat{\kappa}^Q_{\rm eff}$,  values (in GeV/fm) for experimentally favoured
energies.}
\end{table}

The values for strange quark are approximately  energy independent. 
However for charm and bottom quarks we get large values ($\sim6-38$ GeV/fm). 
Earlier analysis~\cite{Magas2002} indicated  
the evidences  of large values for string tensions, $\kappa\sim5-12$ GeV/fm.
The applicability of such  large values can be verified  after performing
proper string model based calculations.  
As it was demonstrated in Ref.~\cite{Pop:2009sd}, 
the available experimental data on charm production at RHIC (for $\sqrt{s}=200$A GeV)
is successfully  described by an 
effective string tension $\sim3$ GeV/fm (the corresponding value in Table II is 
close to this result).

\section{Conclusion}

We have calculated non-perturbative quark pair production in time-dependent strong
non-Abelian SU(2) fields.
Applying a pulse
like time evolution and investigating the influence of pulse duration time,
we observed that light and strange quark-pairs are produced as
we expected, approaching the Schwinger limit.
Charm and bottom  quarks followed this behaviour for long pulses.
However, for short pulses we did not see the expected heavy 
quark suppression, connected to the large  quark mass in the Schwinger estimate.
Indeed, the large value of inverse pulse duration time,
overwhelming the mass of the heavy quark, $1 / \tau \gg m_c$,
determines the quark-pair production.
We obtained enhanced heavy quark production 
at small duration time of the pulse, which can be connected 
to ultrarelativistic  heavy ion collisions.

On the basis of our numerical results obtained from the kinetic equations 
we defined ``flavour specific'' effective string tension values to describe 
the enhanced heavy quark production. The validity of 
the obtained values must be verified by string based Monte-Carlo calculations. 
However our values seem to be reasonable in comparison to previously published string model  
results. 

Finally, we would like to emphasis the strength of our model demonstrated in this paper:
we are able to describe non-perturbative particle production (in strong non-Abelian field)
in a wide energy range, simulating the environment of heavy ion collisions at  different energies. 
The flexibility of our model is very much favoured to understand the experimental data 
and the physics behind them at different collisional energies. 
In the widely used CGC model~\cite{Lappi:2006fp} asymptotic solution could have been extracted 
displaying heavy ion reactions with infinite collisional energy, only. 
Our results, the obtained energy-dependent enhancement of non-perturbative 
heavy quark production, display the complexity of strong field
physics and the importance of continuously varying energy dependence.

\section*{Acknowledgments}
We thank M. Gyulassy for useful comments and  discussions.  
This work was supported in part by  Hungarian OTKA Grants 
NK077816, MTA-JINR Grant, BMBF project RUS 08/038, and 
the RFBR grant No. 08-02-01003-a.

\section*{Appendix A. Kinetic equation for the Wigner function}
The derivation of the kinetic equation for the Wigner function
in non-Abelian case 
was   discussed in details in Ref.~\cite{HeinzOchs}. The 
covariant proper time formulation of the kinetic equation
was extended  from earlier Abelian  version~\cite{Holl02} to
non-Abelian one~\cite{Prozor03}. 
In our investigation, here and earlier~\cite{SkokLev07,Levai:2008wf},  
we use this  non-Abelian version. However,  the original
paper of Prozorkevich et al.~\cite{Prozor03}
contains a few misprints in important equations, which may confuse the reader
and questioning the validity of our work. To avoid this confusion we shortly display 
the most important steps for derivation of the
kinetic equation for the Wigner function
in non-Abelian case.

The starting point is 
the  equations of motion for the quark field operators, 
obtained by the variation of the QCD Lagrangian.
The last reads 
\begin{equation}
{\cal L} = \bar{\psi}( i\gamma^\mu D_\mu - m)\psi - \frac12 F_{\mu \nu} F^{\mu \nu},  
\end{equation}
where $D_\mu= \partial_\mu - ig A_\mu$ is the covariant derivative, $A_\mu = A_\mu^a t^a$
is the 4-potential of color field, $m$ is the current  quark mass and $g$ is the  QCD coupling  
constant. The field tensor $F_{\mu \nu}$ is given by 
\begin{equation}
F_{\mu\nu} = \frac{i}{g} \left[D_\mu , D_\nu \right] = 
\partial_\mu A_\nu - \partial_\nu A_\mu - i g  [ A_\mu, A_\nu ]. 
\end{equation}
The generators of SU(N) group in the fundamental representation $t_a=\lambda_a/2$ 
are expressed in terms of Gell-Mann matrices for SU(3) and Pauli matrices for SU(2). 

The wanted  equations of motion for  spinor field operators read 
\begin{eqnarray}
\partial_t \psi &=& - \gamma^0 \gamma^i D_i \psi - i m \gamma_0 \psi + ig A_0 \psi, \\
\partial_t \bar{\psi} &=&  \bar{\psi} \gamma^0 \gamma^i D_i^*  + i m \bar{\psi} \gamma_0  - ig  \bar{\psi} A_0.  
\end{eqnarray}
Here the covariant derivative $D_i^*$ in the second equation  acts to the left.  

The single-time  Wigner function~\cite{HeinzOchs} is defined by 
\begin{eqnarray}
\label{Wgen}
&&W = \\ 
&&\int d^3y e^{i\mathbf{p} \mathbf{y}} 
U \left(\mathbf{x}, \mathbf{x} + \frac{\mathbf{y}}{2}\right)
\rho (\mathbf{x},\mathbf{y},t) U \left(\mathbf{x}-\frac{\mathbf{y}}{2},\mathbf{x}\right), 
\nonumber
\end{eqnarray}
the unitary link operator $U(\mathbf{x}_1,\mathbf{x}_2)$ is introduced to 
maintain gauge invariance of the Wigner function. The link operator is given by 
\begin{equation}
U({x}_1,{x}_2) = \exp\left\{i g \int_{x_1}^{x_2} dz^\mu A^\mu(z)\right\}.
\end{equation}
The oneparticle density matrix reads 
\begin{equation}
\rho^{ab}_{ik} = - \frac12 \left[  \psi^a_i\left(\mathbf{x}+\frac{\mathbf{y}}{2}\right) , 
\bar{\psi}^b_k\left(\mathbf{x}-\frac{\mathbf{y}}{2}
\right) \right]. 
\end{equation}

We apply  the first derivative w.r.t.  time to both sides of Eq.~(\ref{Wgen})
and take into account that the variation of the link operator is given by 
\begin{eqnarray}
&&\delta U(x_1,x_2) = \nonumber \\ 
&&
\nonumber 
i g \delta x_1^\mu A_\mu(x_1) U(x_1,x_2)
- ig \delta x_2^\mu U(x_1,x_2) A_\mu(x_2)- \\
&& \nonumber 
- ig \int_0^1 ds U(x_1,z) F_{\mu \nu} (z) U(z,x_2) (x_1^\mu - x_2^\mu) \times \\
&& 
\left\{ \delta x_2^\nu + s ( \delta x_1^\nu - \delta x_2^\nu )\right\}, 
\quad z = x_2 + s (x_1-x_2).
\label{varU}
\end{eqnarray}
Defining   the Schwinger string
\begin{equation}
A^{[x_1]}(x_2) = U(x_1, x_2) A(x_2) U(x_2, x_1)
\end{equation}
we rewrite   Eq.~(\ref{varU})
\begin{eqnarray}
\nonumber
&&\delta U(x_1,x_2) = \left\{
i g \delta x_1^\mu A_\mu(x_1) 
- ig \delta x_2^\mu  A^{[x_1]}_\mu(x_2)-  \right.  \\
&& \nonumber 
- ig \int_0^1 ds  F^{[x_1]}_{\mu \nu} (z) (x_1^\mu - x_2^\mu)  
\left. 
\left[ \delta x_2^\nu + s ( \delta x_1^\nu - \delta x_2^\nu )\right]
\right\}  \times \\ 
&&U(x_1,x_2). \nonumber
\end{eqnarray}
Keeping in mind this variation 
we obtain  the wanted  equation for the Wigner function:
\begin{eqnarray}
&&\partial_t W + 
\frac12 \left[ \gamma^0 \gamma^i, \frac{\partial}{\partial x^i} W  \right]+
i m [\gamma_0, W] - i k_i \left\{ \gamma_0 \gamma^i , W \right\} + \nonumber \\
&&\frac{g}{2} \frac{\partial}{\partial k_i}  \int_0^1 ds 
\left\{
E^{[x]}_i(\mathbf{z}^-) W + W E^{[x]}_i(\mathbf{z}^+) + 
\right. \nonumber 
\\ \label{nonlocalW} 
&&\left.
\frac12 F^{[x]}_{i\nu} (\mathbf{z}^-) 
\left( [W,\gamma^0\gamma^\nu] - s \left\{ W, \gamma^0 \gamma^\nu\right\}   \right)
\right. 
\\&&\left.
+ \frac12 \left( [W,\gamma^0\gamma^\nu] + s \left\{ W, \gamma^0 \gamma^\nu\right\}   \right)
F^{[x]}_{i\nu} (\mathbf{z}^+) 
\right\} \nonumber
\\&& \nonumber 
- ig [A_0, W] - \frac12 ig [\gamma_0 \gamma^i,[A_i,W]] =0. 
\end{eqnarray}
Here we used the equation of motions for fermion field operators 
and the following identity   
\begin{equation}
\int f(\mathbf{y}) \exp\left( i \mathbf{k} \mathbf{y} \right) d^3 y = \int
f\left(-i \frac{\partial}{\partial \mathbf{k}} \right)\exp\left( i \mathbf{k} \mathbf{y} \right) d^3 y,
\end{equation}
that is valid  for any analytical function $f(\mathbf{y})$.
The argument of  field tensor is given by
\begin{equation}
\mathbf{z}^\pm = \mathbf{x} \pm  i\frac{s}{2} \frac{\partial}{\partial \mathbf{k}}.
\end{equation}

As we can see in Eq.~(\ref{nonlocalW}), the obtained kinetic equation is nonlocal
and difficult to  solve  numerically. With help of gradient expansion we
can derive  local approximations.
In our calculations  we use  space homogeneous fields. In this case  
the kinetic equation take the form of  Eq.~(\ref{KEW}), that we used in this paper
and previous ones~\cite{SkokLev07,Levai:2008wf}.

The Wigner equation for the vacuum state   follows  from the definition displayed in 
Eq.~(\ref{Wgen}).
Indeed, using usual anticommutation relations for fermion field operators we obtain
the Wigner function in vacuum 
\begin{equation}
W^{ab}_{\rm v} = - \frac{m+\mathbf{k}\mathbf{\gamma}}{2\omega} \delta^{ab}.
\end{equation}

\section*{Appendix B. Exact solutions for SU(2)-color case}

In Section III
the kinetic equation for the Wigner function was 
solved numerically. However, 	   
after taking into account 
additional symmetries of the external field 
we can discover  further simplifications
and even obtain    exact solutions.

To demonstrate this fact we rewrite  Eqs. (\ref{final_bs}-\ref{final_em})
explicitly for the distribution function  $f(\mathbf{k},t)$:
\begin{eqnarray}
\label{f_beg}
&& \p_t f  + \frac{3}{4} gE^\diamond_z \ppkz  f^\diamond = \frac{3}{4} W^\diamond v^\diamond, \\ 
&& \p_t v^s  + \frac{3}{4} gE^\diamond_z \ppkz  v^\diamond = -\frac{3}{4} W^\diamond f^\diamond - 2 \omega u^s, \\ 
&& \p_t u^s  + \frac{3}{4} gE^\diamond_z \ppkz  u^\diamond = 2 \omega v^s, \\ 
&& \p_t f^\diamond  +  gE^\diamond_z \ppkz  f =  W^\diamond v^s, \\ 
&& \p_t v^\diamond  +  gE^\diamond_z \ppkz  v^s = \frac{1}{2} W^\diamond (1-2f) - 2 \omega u^\diamond, \\ 
&& \p_t u^\diamond  + gE^\diamond_z \ppkz  u^s = 2 \omega v^\diamond. 
\label{f_end}
\end{eqnarray}
Here the following new functions were defined  
\begin{eqnarray}
f^\diamond &=& \frac{m a^\diamond + \mathbf{k}\mathbf{b}^\diamond}{\omega}, \\
v^{s,\diamond} &=&\frac{\varepsilon_\perp}{\omega} b_z^{s,\diamond} - \frac{k_z}{\omega}\left(\frac{ma^{s,\diamond} 
+ k_\perp b^{s,\diamond}_\perp } {\varepsilon_\perp}\right), \\
u^{s,\diamond} &=& - \frac{k_\perp}{\varepsilon_\perp}d^{s,\diamond}  - 2 \frac{m}{\varepsilon_\perp} c_z^{s,\diamond}, \\
W^\diamond &=& \frac{gE \varepsilon_\perp}{\omega^2}.
\label{W}
\end{eqnarray}
The
naming scheme is chosen to be  consistent with the U(1)
case of our previous work~\cite{SkokLev05}.
The transverse one-particle energy in Eq.~(\ref{W}) is defined by 
$\varepsilon_\perp=\sqrt{k_\perp^2+m^2}$.

As it follows from
Eq.~(\ref{Initial}) the vacuum state corresponds to   zero initial conditions for the functions
$f, f^{\diamond}$, $v^{s,\diamond}$, $u^{s,\diamond}$. 

Note, that Eqs.~(\ref{f_beg}-\ref{f_end}) has the same number of equation for massive 
and massless particles. Thus the massless limit does not lead to any further simplifications.   
The system (\ref{f_beg}-\ref{f_end}) can be transformed to more conventional form, allowing 
the solution on characteristics. For that we introduce the following new  functions  
\begin{eqnarray}
F^{\pm} &=& f \pm \frac{\sqrt{3}}{2} f^\diamond, \\
V^{\pm} &=& v^s \pm \frac{\sqrt{3}}{2} v^\diamond, \\
U^{\pm} &=& u^s \pm \frac{\sqrt{3}}{2} u^\diamond. 
\end{eqnarray}
The equations for these functions read
\begin{eqnarray}
\label{Fb}
\p_t F^{\pm}  \pm \frac{\sqrt{3}}{2} gE^\diamond_z \ppkz  F^{\pm} &=&  \pm \frac{\sqrt{3}}{2} W^\diamond V^{\pm}, \\ 
\p_t V^{\pm}  \pm \frac{\sqrt{3}}{2} gE^\diamond_z \ppkz  V^{\pm} &=& \pm \frac{\sqrt{3}W^\diamond}{4}(1-2F^\pm) - \nonumber \\
&&-2 \omega U^\pm, \\ 
\p_t U^{\pm}  \pm \frac{\sqrt{3}}{2} gE^\diamond_z \ppkz  U^{\pm} &=& 2 \omega V^\pm.
\label{Fe}
\end{eqnarray}
The equations for ``(+)'' and  ``(--)''   functions are  completely factorized and can be solved independently. The  distribution function $f$ is obtained as 
\begin{equation}
f = \frac{F^++F^-}2.
\end{equation}
The equations (\ref{Fb}-\ref{Fe}) are very similar to those  obtained in the Abelian case  
(see  Eqs. (76-77) in \cite{Pervushin:2006vh} and Eqs. (22-24) in \cite{SkokLev05}). 
However, only $f$ is physical quantity, while the  functions $F^\pm$ carry  intermediate
information.     
Nevertheless, from mathematical point of 
view there is no difference and  this analogy allows to  exploit U(1) solution 
to obtain exact analytical results for SU(2)-color case, as we demonstrate  below.   

In the case  of a  time reversal symmetry of the field 
strength, $E(t)=E(-t)$,
the corresponding components are equal to  each other,
e.g. $F^{+}=F^{-}$. Thus 
the distribution function is defined as $f = F^{+}=F^{-}$. 
Furthermore in the Abelian case 
it is known that for the field with form of Eq.~(\ref{field})  
an analytic solution of Dirac equation  exists,  
as well as  analytic solution of the  corresponding kinetic equation (see e.g. Ref.~\cite{grib94}). 
Following this analogy 
we can obtain  an  analytic solution  
of Eqs. (\ref{Fb}-\ref{Fe}) in the asymptotic  state~$t\gg\tau$:
\begin{equation}
\label{ansol}
f = \frac{\sinh\left(\pi (\theta - \mu^+ +\mu^-)  \right) \sinh\left(\pi (\theta + \mu^+ -\mu^-)  \right)     }
{\sinh(2\pi\mu^+) \sinh(2\pi\mu^-)}.\quad \quad 
\end{equation}
Here we introduced the following notations 
\begin{eqnarray}
&&\mu^\pm = \frac{\tau}{2} 
\sqrt{\left(k_z \pm  \frac{E_0}{E_{cr}} m^2 \tau \right)^2+k_\perp^2 + m^2 },\\ 
&&\theta = - \frac{E_0}{E_{cr}} m^2 \tau^2, \\
&& E_{cr} = \frac{2 m^2}{\sqrt{3} g}.
\end{eqnarray}

The {\it Schwinger limit} can be  readily obtained from
Eq.~(\ref{ansol}). Indeed,
for pulses longer than any scale in the 
system, 
$\tau\gg {\rm max}\{E_0^{-1/2}, m^{-1/2}\}$, 
we can use the  expansions 
\begin{eqnarray}
&&\theta \pm ( \mu^+ - \mu^- ) = -\frac{E_0}{E_{cr}} m^2\tau^2 
\mp k_3 \tau + {\cal O}\left(\frac1{\tau}\right), \\   
&&\mu^\pm = \frac{E_0}{2 E_{cr}} m^2 \tau^2 \pm k_3 \tau + 
\frac{2}{\sqrt{3}} \frac{m^2+k_\perp^2}{4 g E_0}  + {\cal O}\left(\frac1{\tau}\right).\ \  \quad 
\end{eqnarray}
In the leading order of expansion parameter $\tau$ 
we obtain  the following  distribution
function  
\begin{equation}
f \simeq \exp\left(-\pi \frac{2 \varepsilon_\perp}{\sqrt{3} g E_0}\right). 
\end{equation}
After the integration w.r.t.  momentum we obtain SU(2) version of the Schwinger formula
\begin{equation}
n = \tau  \frac{3 (g E_0)^2}{16 \pi^2} \exp\left(  - \frac{2\pi m^2}{\sqrt{3}g E_0} \right), 
\end{equation}
where we have taken into account the replacement $\int d k_z \to \sqrt{3} gE_0 \tau /2 $.
Thus in SU(2) the suppression factor of  heavy particles with mass $m_Q$ to
light particles with mass $m_q$  in Schwinger limit is given by
\begin{equation}
\gamma^Q = \exp\left(  - \frac{2\pi (m_Q^2-m_q^2)}{\sqrt{3}g E_0} \right). 
\end{equation}

The analytical result for the number density of quarks 
can be also obtained if the following inequality 
is satisfied 
\begin{equation}
\Gamma_K\equiv  \frac{E_{cr}}{E} \frac1{m \tau}  \gg {\rm max} \{ 1, m\tau\}, 
\label{gammak}
\end{equation}
where we introduce the adiabaticity parameter $\Gamma_K$~\footnote{In the Abelian case 
$\Gamma_{K}$ is known as the Keldysh adiabaticity parameter. 
It separates the nonperturbative region $\Gamma_{K}\ll1$ from 
the perturbative multiphoton one $\Gamma_{K}\gg1$.}.
In this limit  the dictribution function is given by 
\begin{equation}
f \simeq \left(  \frac{ \sqrt{3} \pi \sqrt{k_\perp^2+m^2 } } {2 \omega} g E \tau^2 
{\rm csch} \left( \pi \omega \tau \right) \right)^2.
\label{approxf}
\end{equation}
The  momentum integration of this distribution function can be done 
analytically  in the following cases:
\begin{itemize}
\item[a)] {\em Long pulse duration and undercritical field.}
For long pulse duration 
the constraint  $m\tau\gg1$ in Eq.~(\ref{gammak}) is satisfied for  
undercritical field,  $E/E_{cr}\ll1$.
After expanding the distribution function 
in Eq.~(\ref{approxf}) and performing momentum integration, 
we obtain the number density  
\begin{equation}
n = \frac{3 (m\tau)^{3/2}}{4\pi} \tau (gE_0)^2 \exp( - 2 \pi m \tau).
\end{equation}

The condition of long pulse duration and 
undercritical field 
could be realized for the collision of heavy ions
in SPS energy range (assuming that the physical picture 
of classical gluon field is still valid) for light quarks, or 
for charm and bottom quarks at and below the RHIC energies 
(the string tension is the order of  $1$ GeV/fm.) 

\item[b)] {\em Short pulse duration. }
In the  opposite limit, $m\tau\ll1$,  we obtain the number density   
\begin{equation}
n \simeq \frac{\tau}{12\pi} (g E_0)^2.
\label{shortpulse}
\end{equation}
In this case   
the number density of produced particles depends  linearly on the  duration
time, $\tau$, and is independent of  the particle  mass, $m$.

For the parameter set  we used in the main part of the manuscript the field 
magnitude, $E_0$, is about five times  higher than  the critical one for the strange 
quark. 
The number of produced strange
quarks follows the Eq.~(\ref{shortpulse})
for pulse duration at least five  times less  
then inverse mass of strange quark 
$ \tau \ll (5 m_s)^{-1}\simeq 0.3$ fm$/c$
As we estimated in  Eq.~(\ref{taus}) the pulse duration time for 
RHIC is about  $0.1$ fm/c, that is only three times less than 
$(5 m_s)^{-1}$. Thus 
the expression (\ref{shortpulse}) is valid for the strange quark
only at higher than RHIC  energies. 

For heavier particles, 
e.g. charm quark, the requirement  $m_c \tau \ll 1$
leads to smaller values of  $\tau$. 
Indeed, we can rewrite 
\begin{equation}
m_c \tau = \frac{m_c}{m_s} \times (m_s \tau) \ll 1.  
\end{equation}
This  expression shows that Eq.~(\ref{shortpulse}) 
becomes 
valid for charm quark on a shorter scale of pulse duration time ($m_c/m_s \simeq 8$).

For light quarks the condition  $m_u \tau \ll 1$
is trivially satisfied at  RHIC energies.
However now the condition~(\ref{gammak}) 
plays more important role.  From Eq.~(\ref{gammak}) 
we obtain an estimate for the validity 
of  Eq.~(\ref{shortpulse})
\begin{equation}
\tau\ll 4\cdot 10^{-3} {\rm fm}/c.  
\end{equation}
In this limiting case the strange suppression factor tends to unity, 
that supports our numerical result in Fig.~\ref{gamma}..

\end{itemize}

One more  analytical result can be obtained from  the general solution 
Eq.~(\ref{ansol}). 
If the duration time of the pulse  tends to zero, but amplitude  
is increasing as $E_0=A_0\tau^{-1}$ ($A_0$ is constant), 
then the distribution function  is given by 
\begin{equation}
f = \frac12\left(1  -  \frac{\omega^2-\frac34 A_0^2}{\omega_+ \omega_-  }\right). 
\label{fdiverg}
\end{equation}
where $\omega_{\pm} = 2 \mu_\pm / \tau$.
Since the momentum integral from the above  distribution function
is divergent, this approximation  results in  infinitely many new
quark-antiquark pairs  in unit volume. This is understandable since we pump 
infinite energy to the system in this special case.


\begin{thebibliography}{99}


\bibitem{schw51} J. Schwinger, Phys. Rev. 82 (1951) 664.


\bibitem{Adams:2004fc}
J.~Adams {\it et al.}  [STAR Collaboration],
Phys.\ Rev.\ Lett.\  {\bf 94}, 062301 (2005)
[arXiv:nucl-ex/0407006].

\bibitem{Baumgart:2008xe}
S.~L.~Baumgart  [STAR Collaboration],
arXiv:0805.4228 [nucl-ex].

\bibitem{Abelev:2008hja}
B.~I.~Abelev {\it et al.}  [STAR Collaboration],
arXiv:0805.0364 [nucl-ex].

\bibitem{Adare:2006hc}
A.~Adare {\it et al.}  [PHENIX Collaboration],
Phys.\ Rev.\ Lett.\  {\bf 97}, 252002 (2006)
[arXiv:hep-ex/0609010].
\bibitem{Adare:2006nq}
A.~Adare {\it et al.}  [PHENIX Collaboration],
Phys.\ Rev.\ Lett.\  {\bf 98}, 172301 (2007)
[arXiv:nucl-ex/0611018].




\bibitem{QM08}
Quark Matter'08 Conference Proceedings
(Ed. by Feng Liu, Zhigang Xiao, Pengfei Zhuang),
J. Phys. {\bf G 36} (2009).




\bibitem{Frawley:2008kk}
A.~D.~Frawley, T.~Ullrich, and R.~Vogt,
Phys.\ Rept.\  {\bf 462}, 125 (2008)
[arXiv:0806.1013 [nucl-ex]].


\bibitem{Pop:2009sd}
V.~Topor Pop, J.~Barrette, and M.~Gyulassy,
Phys.\ Rev.\ Lett.\  {\bf 102}, 232302 (2009)
[arXiv:0902.4028 [hep-ph]].





\bibitem{FRITIOF} 
B. Andersson {\it et al.}, 
Phys. Rep. {\bf 97} (1983) 31;
Nucl. Phys. {\bf B281} (1987) 289. 

\bibitem{HIJ}
X.N. Wang and M. Gyulassy,
Phys. Rev. {\bf D44} (1991) 3501;
Comput. Phys. Commun. {\bf 83} (1994) 307.

\bibitem{RQMD}
H. Sorge,
Phys. Rev. {\bf C52} (1995) 3291.

\bibitem{QGSM}
N.~S.~Amelin, K.~K.~Gudima, S.~Y.~Sivoklokov, and V.~D.~Toneev,
Sov.\ J.\ Nucl.\ Phys.\  {\bf 52}, 172 (1990)
[Yad.\ Fiz.\  {\bf 52}, 272 (1990)].

\bibitem{HIJINGBB}
V.~Topor Pop, M.~Gyulassy, J.~Barrette, C.~Gale, X.~N.~Wang, and N.~Xu,
Phys.\ Rev.\  C {\bf 70}, 064906 (2004)
[arXiv:nucl-th/0407095].


\bibitem{Gyul97}
M. Gyulassy and L. McLerran,
Phys. Rev. {\bf C56} (1997) 2219.

\bibitem{Topor05}
V.~Topor Pop, M.~Gyulassy, J.~Barrette, C.~Gale, R.~Bellwied, and N.~Xu,
Phys.\ Rev.\  C {\bf 72}, 054901 (2005); \\
V.~Topor Pop, M.~Gyulassy, J.~Barrette, C.~Gale, S.~Jeon, and R.~Bellwied,
arXiv:hep-ph/0608136; \\
Phys.\ Rev.\  C {\bf 75}, 014904 (2007).



\bibitem{Lappi:2006fp}
T.~Lappi and L.~McLerran,
Nucl.\ Phys.\  A {\bf 772}, 200 (2006)
[arXiv:hep-ph/0602189].







\bibitem{McLerran:2001sr}
L.~D.~McLerran,
Lect.\ Notes Phys.\  {\bf 583}, 291 (2002)
[arXiv:hep-ph/0104285].



\bibitem{Magas2002}
V.~K.~Magas, L.~P.~Csernai, and D.~Strottman,
Nucl.\ Phys.\  A {\bf 712}, 167 (2002)
[arXiv:hep-ph/0202085].




\bibitem{IanVen03}
E. Iancu and R. Venugopalan,
{\tt hep-ph/0303204} and references therein.



\bibitem{Gelis:2006yv}
F.~Gelis and R.~Venugopalan,
Nucl.\ Phys.\  A {\bf 776}, 135 (2006)
[arXiv:hep-ph/0601209].


\bibitem{Gelis:2006cr}
F.~Gelis and R.~Venugopalan,
Nucl.\ Phys.\  A {\bf 779}, 177 (2006)
[arXiv:hep-ph/0605246].


\bibitem{Blaizot:2004wv}
J.~P.~Blaizot, F.~Gelis, and R.~Venugopalan,
Nucl.\ Phys.\  A {\bf 743}, 57 (2004)
[arXiv:hep-ph/0402257].


\bibitem{Gatoff87}
G. Gatoff, {\it et al.} 
Phys. Rev. {\bf D36} (1987) 114.

\bibitem{Kluger91}
Y. Kluger {\it et al.},
Phys. Rev. Lett. {\bf 67} (1991) 2427.

\bibitem{Gatoff92}
G. Gatoff and C.Y. Wong, Phys. Rev. {\bf D46} (1992) 997.

\bibitem{Wong95}
C.Y. Wong, {\it et al.}
Phys. Rev. {\bf D51} (1995) 3940.



\bibitem{Eisenberg95}
J.M. Eisenberg, Phys. Rev. {\bf D51} (1995) 1938.

\bibitem{Vin02}
D.V. Vinnik, {\it et al.},
Few-Body Syst. {\bf 32} (2002) 23 .

\bibitem{Per03}
V.N. Pervushin, {\it et al.}
accepted in Int. Mod. Phys. A. ({\tt hep-th/0307200}).



\bibitem{Prozorkevich:2004yp}
A.~V.~Prozorkevich, {\it et al.} 
Phys.\ Lett.\  B {\bf 583} (2004) 103
[arXiv:nucl-th/0401056].


\bibitem{Pervushin:2006vh}
V.~N.~Pervushin and V.~V.~Skokov,
Acta Phys.\ Polon.\  B {\bf 37}, 2587 (2006)
[arXiv:astro-ph/0611780].



\bibitem{Mihaila:2009ge}
B.~Mihaila, F.~Cooper, and J.~F.~Dawson,
arXiv:0905.1360 [hep-ph].

\bibitem{Dawson:2009cn}
J.~F.~Dawson, B.~Mihaila, and F.~Cooper,
arXiv:0906.2225 [hep-ph].



\bibitem{Prozor03}
A.V. Prozorkevich, S.A. Smolyansky, and S.V. Ilyin,
({\tt hep-ph/0301169}).

\bibitem{HeinzOchs}
H.~T.~Elze, M.~Gyulassy, and D.~Vasak,
Phys.\ Lett.\  B {\bf 177}, 402 (1986);
H.~T.~Elze, M.~Gyulassy, and D.~Vasak,
Nucl.\ Phys.\  B {\bf 276}, 706 (1986);
S. Ochs and U. Heinz,
Ann. Phys. {\bf 266} (1998) 351.




\bibitem{Diet03}
D.D. Dietrich,
Phys. Rev. {\bf D68} (2003) 105005;
{\it ibid.} {\bf D70} (2004) 105009.





\bibitem{SkokLev05}
V.V. Skokov and P. Levai,
Phys. Rev. {\bf D71} (2005), 094010, ({\tt hep-ph/0410339}).

\bibitem{SkokLev07}
V.~V.~Skokov and P.~Levai,
Phys. Rev. {\bf D78} (2008) 054004,
({\tt arXiv:0710.0229 [hep-ph]}).

\bibitem{Levai:2008wf}
P.~Levai and V.~Skokov,
J. Phys. G: Nucl. Part. Phys. {\bf 36} (2009) 064068
[arXiv:0812.2536].



\bibitem{Levai:2000ne}
P.~Levai, T.~S.~Biro, P.~Csizmadia, T.~Csorgo and J.~Zimanyi,
J.\ Phys.\ G {\bf 27}, 703 (2001)
[arXiv:nucl-th/0011023].

\bibitem{Levai:2008me}
P.~Levai,
J.\ Phys.\ G {\bf 35}, 044041 (2008)
[arXiv:0806.0133 [nucl-th]].

\bibitem{Gyulassy:1985}
M.~Gyulassy and A.~Iwazaki,
Phys.\ Lett.\  {\bf 165B}, 157 (1985).

\bibitem{Holl02}
A. H\"oll, V.G. Morozov, and G. R\"opke,
Ther. Math. Phys. {\bf 131}, 812 (2002); 
Ther. Math. Phys. {\bf 132}, 1029 (2002). 







\bibitem{grib94}
A.A. Grib, S.G. Mamaev, and V.M. Mostepanenko,
{\it Vacuum Quantum Effects in Strong Fields}, 
(Friedmann Laboratory Publishing, St. Petersburg, 1994); 
G.~V.~Dunne,
arXiv:hep-th/0406216;
K.~Fukushima, F.~Gelis and T.~Lappi,
arXiv:0907.4793 [hep-ph].



\end{thebibliography}
\end{document}